\documentclass[aps,floatfix,superscriptaddress,twocolumn,tightenlines,amsmath,amssymb,nofootinbib,10pt,longbibliography,prl]{revtex4-1}

\usepackage[utf8]{inputenc}
\usepackage[T1]{fontenc}
\usepackage[mathscr]{euscript}
\usepackage{graphicx}
\usepackage[dvipsnames]{xcolor}
\usepackage{amsthm}
\usepackage{psfrag}
\usepackage{ifsym}
\usepackage{dsfont}
\usepackage{enumerate}
\usepackage{amsfonts}
\usepackage{multirow} 
\usepackage{amsmath}
\usepackage{hyperref}
\hypersetup{colorlinks=true,linkcolor=red,citecolor=blue}
\usepackage[sort&compress]{natbib}
\usepackage[separate-uncertainty=true]{siunitx}
\usepackage{comment}
\usepackage{ulem}

\newcommand{\bra}[1] {\langle #1 |}
\newcommand{\ket}[1] {| #1 \rangle}

\newcommand{\Tr} {\operatorname{Tr}}
\newcommand{\expec}[1]{\left\langle #1 \right\rangle}

\newcommand\blfootnote[1]{%
  \begingroup
  \renewcommand\thefootnote{}\footnote{#1}%
  \addtocounter{footnote}{-1}%
  \endgroup
}

\begin{document}

\title{Experimental quantum conference key agreement}

\author{Massimiliano Proietti$^{\dagger}$}

\blfootnote{$^{\dagger}$ These two authors contributed equally.} 
\affiliation{Institute of Photonics and Quantum Sciences, School of Engineering and Physical Sciences, Heriot-Watt University, Edinburgh EH14 4AS, UK}

\author{Joseph Ho$^{\dagger}$}

\affiliation{Institute of Photonics and Quantum Sciences, School of Engineering and Physical Sciences, Heriot-Watt University, Edinburgh EH14 4AS, UK}

\author{Federico Grasselli}
\affiliation{Institut für Theoretische Physik III, Heinrich-Heine-Universität Düsseldorf, Universitätsstraße 1, D-40225 Düsseldorf, Germany}

\author{Peter Barrow}
\affiliation{Institute of Photonics and Quantum Sciences, School of Engineering and Physical Sciences, Heriot-Watt University, Edinburgh EH14 4AS, UK}

\author{Mehul Malik}
\affiliation{Institute of Photonics and Quantum Sciences, School of Engineering and Physical Sciences, Heriot-Watt University, Edinburgh EH14 4AS, UK}

\author{Alessandro Fedrizzi}
\affiliation{Institute of Photonics and Quantum Sciences, School of Engineering and Physical Sciences, Heriot-Watt University, Edinburgh EH14 4AS, UK}

\maketitle

\textbf{Quantum networks will provide multi-node entanglement over long distances to enable secure communication on a global scale.
Traditional quantum communication protocols consume pair-wise entanglement, which is sub-optimal for distributed tasks involving more than two users.
Here we demonstrate quantum conference key agreement, a quantum communication protocol that exploits multi-partite entanglement to efficiently create identical keys between N users with up to N-1 rate advantage in constrained networks.
We distribute four-photon Greenberger-Horne-Zeilinger (GHZ) states generated by high-brightness, telecom photon-pair sources across up to 50~km of fibre, implementing multi-user error correction and privacy amplification on resulting raw keys.
Under finite-key analysis, we establish \(1.15\times10^6\) bits of secure key, which are used to encrypt and securely share an image between the four users in a conference transmission.
We have demonstrated a new protocol tailored for multi-node networks leveraging low-noise, long-distance transmission of GHZ states that will pave the way forward for future multiparty quantum information processing applications.
}

\section*{Introduction}

Conference key agreement (CKA) is a multi-user protocol for sharing a common information-theoretic secure key beyond the two-party paradigm~\cite{CKA}.
This key allows group-wide encryption for authenticated users to communicate securely, wherein exclusively members of the group can decrypt messages broadcast by any other member.
Traditional two-party quantum key distribution (2QKD) primitives~\cite{Peev2009networkSECOQC,Sasaki2011tokyo,dynes2019cambridge,Wengerowsky2018_2QKD} can be used to share N-1 individual key pairs between \(N\) users followed by classical computational steps to distill a conference key.
However, this is inefficient for producing conference keys when users have access to a fully connected quantum network, as envisioned in the `quantum internet'~\cite{kimble2008quantum,wehner2018quantum}. 
An efficient alternative is to derive conference keys directly from multi-partite entangled states created in such networks ~\cite{epping2017multi,grasselli2018finite,MDI-QKDprl}---we refer to these methods as quantum CKA.

Quantum CKA (QCKA) is a generalisation of entanglement-based QKD to $N$ users~\cite{CKA}. The currently most practical QCKA variant is based on the distribution of GHZ states~\cite{grasselli2018finite}. This protocol has been proven secure including for the finite key scenario and offers performance advantages over conference key generation from pair-wise keys (2QKD) under different noise models, channel capacity constraints and network router configurations~\cite{epping2017multi,epping2016networks,ribeiro2018fully,grasselli2018finite,Pivoluska:2018bp,jo19semi,hahn2019networks}. The clearest advantage of QCKA arises in true quantum networks~\cite{epping2016networks2}: GHZ states can be distilled from an underlying network graph state in as little as a single network use while 2QKD requires up to $N-1$ copies to generate the required key pairs~\cite{epping2017multi}.

\begin{figure*}[t!]
 \begin{center}
 \includegraphics[width=178mm]{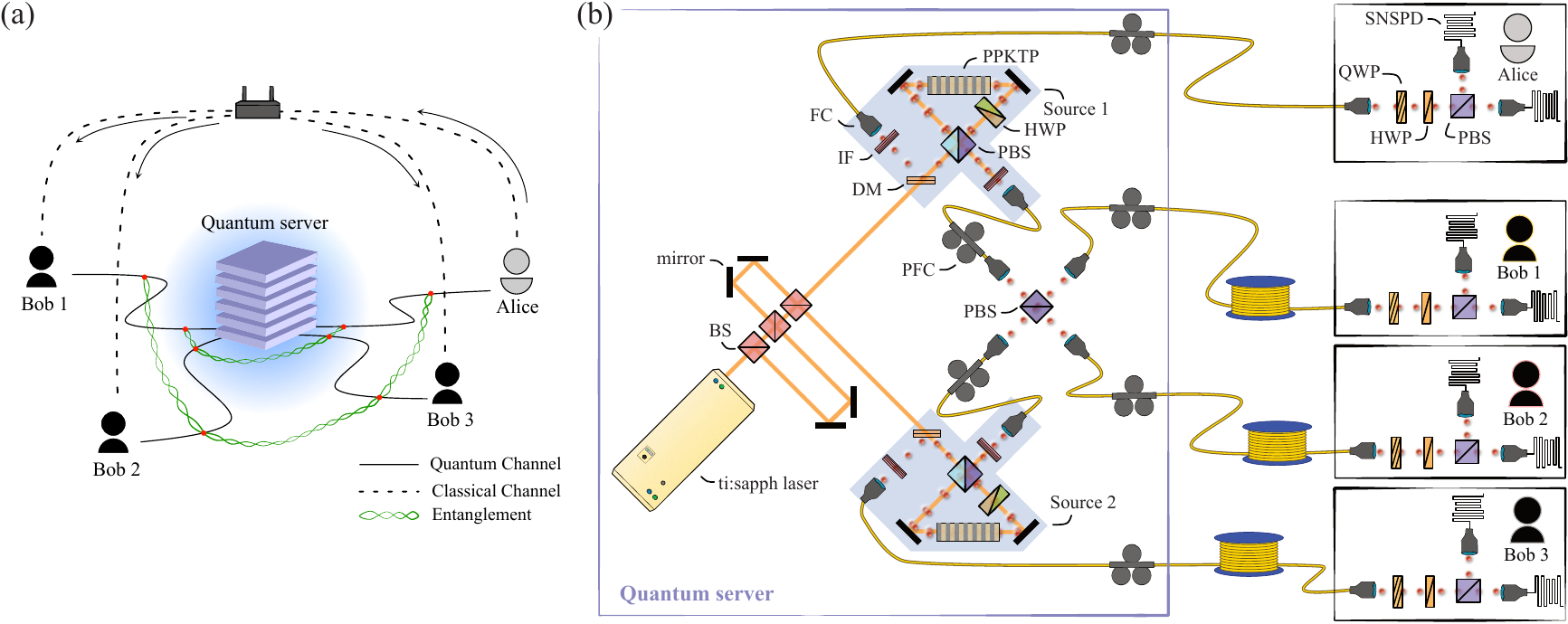}
 \end{center}
 \vspace{-2em}
\caption{\textbf{(a)} Quantum conference key agreement scheme.
A quantum server distributes entangled GHZ states to Alice, who initiates the protocol, and Bobs 1, 2, and 3. They establish a common key from a pre-agreed sequence of Z measurements while checking the security by measuring X.
\textbf{(b)} A mode-locked picosecond laser (ti:sapph) multiplexed to \SI{320}{MHz} repetition rate supplies two entangled photon sources which are based on parametric downconversion in periodically poled KTP crystals (PPKTP), pumped bidirectionally in a Sagnac loop for producing polarisation-entangled Bell pairs~\cite{Fedrizzi2007Sagnac}. Down-converted photons are separated from the pump with dichroic mirrors (DM) and coupled into fibres (FC). One photon from each source non-classically interfere on a polarising beamsplitter (PBS) creating the four-photon GHZ state, see Materials and Methods for details. Each user receives their photon via single-mode fibres and performs projective measurements in the Z(X) basis by using a quarter- (QWP) and half-wave plate (HWP), and a polarising beamsplitter (PBS) before detection with superconducting nanowire single-photon detectors (SNSPD). Detection events are time-tagged and counted in coincidence within a \SI{1}{ns} time window.}
 \label{fig:Figure1}
\end{figure*}

Here we experimentally demonstrate the salient features of the N-BB84 protocol introduced in~\cite{grasselli2018finite} with a state-of-the-art photonic platform.
An untrusted quantum server prepares and distributes $L$ rounds of the maximally entangled GHZ state, 
\( \ket{GHZ} \equiv (\ket{0}^{\otimes N} + \ket{1}^{\otimes N}) /\sqrt{2}\)
, to \(N\) participants in the network.
In our work we implement a four-party protocol consisting of: Alice (A), Bob 1 (\(\text{B}_1\)), Bob 2 (\(\text{B}_2\)), and Bob 3 (\(\text{B}_3\)), see Fig.~\ref{fig:Figure1}~(a).
Each user performs quantum measurements on their respective photon in either the Z-basis \(\{ \ket{0}, \ket{1} \}\) constituting type-1 rounds, or the X-basis \(\{ \ket{+}\doteq(\ket{0}+\ket{1})/\sqrt{2}, \ket{-}\doteq(\ket{0}-\ket{1})/\sqrt{2}\}\) for type-2 rounds.
Type-1 rounds contribute to the raw key as these measurements ensure all users in the protocol obtain the same bit value owing to the structure of the GHZ state.
A small portion of these outcomes will be consumed to determine the error rates.
Type-2 rounds are carried out randomly with probability \(p\), for a total of \(m = L \cdot p\) rounds, and are used to detect the presence of an eavesdropper.
Users coordinate the measurement sequence using \(L \cdot h \left( p \right)\) bits of a pre-shared key. 
In particular, one user generates the \(L\)-bit string indicating the measurement type of each round.
The string can be classically compressed, shared, and decompressed by the other parties.
Note that the values of $p$ are typically on the order of 0.02, leading to a small value of $h(p)$, i.e., the amount of information to be initially pre-shared is small.

Once the measurements are complete, the users proceed to verify the security of their key by performing parameter estimation. 
All users announce their outcomes for a subset of the type-1 rounds, \(m\) in total and randomly chosen, and all \(m\) type-2 rounds to determine \(Q^m_{\text{AB}_i} = \left(1-\expec{\sigma_z^{\text{A}}\sigma_z^{\text{B}_i}}\right)/2\) for \(i=\{1,2,3\}\) and \(Q^m_X = \left(1-\expec{\sigma_x^{\otimes 4}}\right)/2\) respectively.
We define the quantum bit error rate (QBER) as \(\textrm{QBER}^m\doteq\textrm{max } Q_{\text{AB}_i}^m\).
All users retain \(n = L - 2m\) bits forming the raw conference key, subsequently corrected with an error correction scheme and shortened with privacy amplification to ensure security.
Finally, all users remove \(L \cdot h \left( p \right)\) bits from their secret conference key to encode the pre-shared keys for subsequent protocols.
Hence, our protocol is a key-growing routine, as in any known QKD scheme.

\section*{Results}

In our experiment, see Fig.~\ref{fig:Figure1}(b), we employ two high-brightness, polarisation-entangled photon-pair sources~\cite{Fedrizzi2007Sagnac} at telecommunication wavelength (\SI{1550}{nm}).
We generate four-photon GHZ states by non-classically interfering one photon from each source on a PBS, which has success probability of $1/2$ (see for example~\cite{ProiettiGHZ} or Materials and Methods for details).
We use commercially available superconducting nanowire single-photon detectors (SNSPDs) with typical quantum efficiencies of $>80\%$ at this wavelength.

\begin{figure}[t!]
  \begin{center}
  \includegraphics[width=86mm]{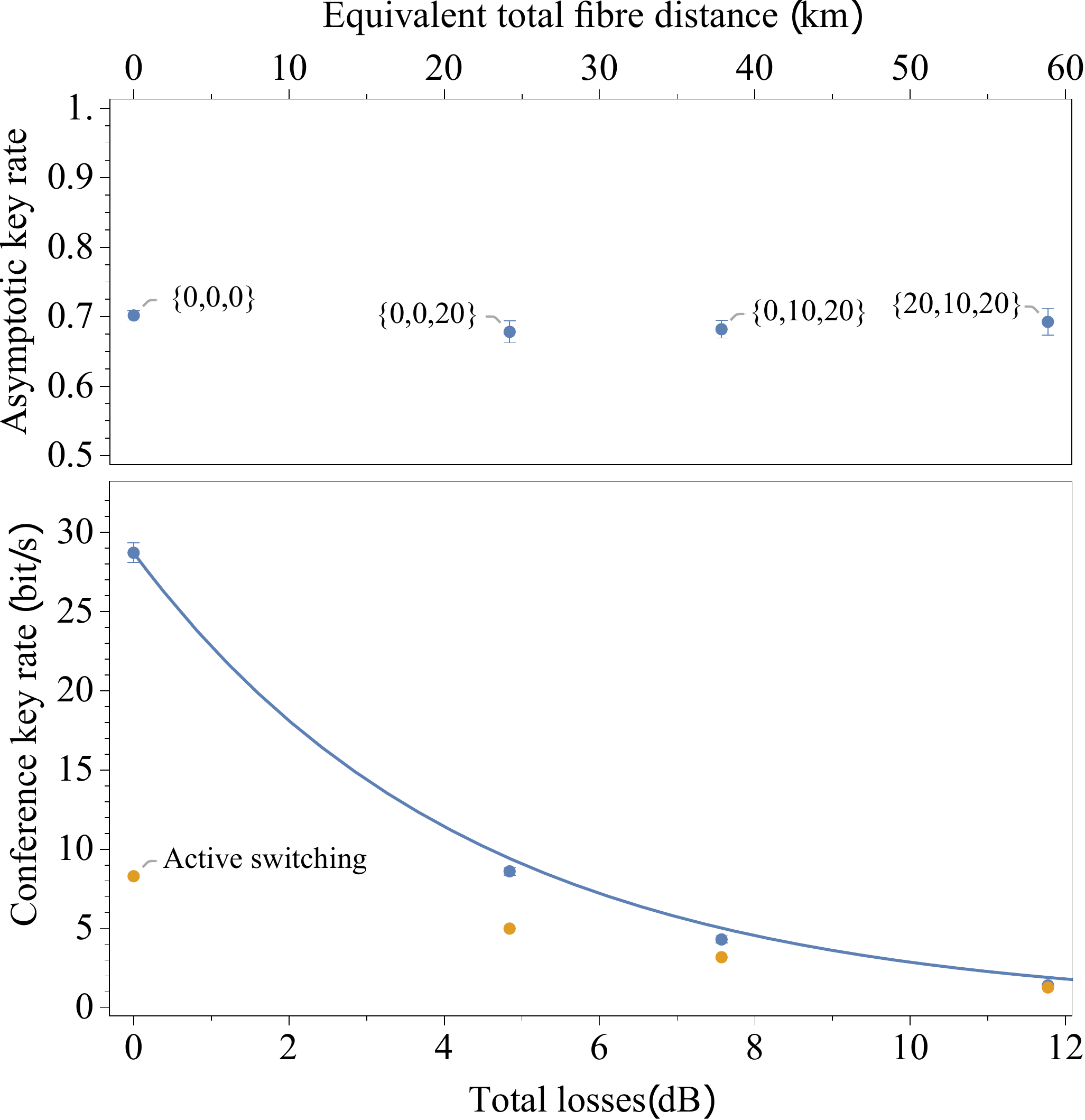}
  \end{center}
  \vspace{-2em}
  \caption{\textbf{Asymptotic key rate results}. (top) We determine the asymptotic key rate (AKR) in Eq.~\ref{AKReq} directly from parameter estimation, i.e., from \(Q_X\) and QBER, and assume theoretical performance of error correction and privacy amplification.
We evaluate AKR for a range of loss conditions set by the placement of fibre links in the network.
(bottom) The conference key rate is plotted as a function of the total fibre length in the network.
We include results of the generation rates with measurement-basis switching using our implementation, see Materials and Methods for details.}
  \label{fig:Figure2}
\end{figure}

We establish the upper bound on the performance of our protocol by assuming an infinite number of rounds can be performed, \(L\rightarrow \infty\).
In this asymptotic regime nearly all rounds are used to extract the raw key, $p\rightarrow0$. 
We evaluate the asymptotic key rate (AKR) as the fraction of secret bits, \(\ell\), extracted from the total rounds~\cite{grasselli2018finite}:
\begin{equation}
    \text{AKR} = \frac{\ell}{L} = 1 - h(Q_X) -  h(\text{QBER}) \,,
    \label{AKReq}
\end{equation}
where $h(x)=-x\log_2 x - (1-x)\log_2 (1-x)$ is the Shannon entropy.
From Eq.~\ref{AKReq} we note the AKR depends only on the noise parameters $Q_X$ and QBER.
We estimate these parameters experimentally using a large sample size of type-1 and type-2 measurements to minimise uncertainties. The results are shown in Fig.~\ref{fig:Figure2}.

We denote the network topology as $\{d_1,d_2,d_3\}$, where $d_i$ is the fibre length in kilometres between $\text{B}_i$ and the server.
Alice remains fixed at \SI{2}{m} from the server in all cases.
We implement four scenarios: $\{0,0,0\}$, $\{0,0,20\}$, $\{0,10,20\}$, and $\{20,10,20\}$, corresponding to measured network losses (in dB) of $0$, $4.84$, $7.57$, and $11.77$.
The observed four-photon generation rates $g_R$ for these scenarios are \SI{40.89}{Hz}, \SI{12.68}{Hz}, \SI{6.31}{Hz}, and \SI{2.03}{Hz}.
The conference key rate is determined as a product of the fractional AKR and the recorded generation rates \(g_R\).
In all cases we observe similar noise parameters, and thus AKR, indicating that the entanglement quality is not degraded significantly by the transmission in fibres.
The experimental AKR is mainly limited by multiple-pair generations at the sources and by spectral impurities of the photons, see Supplementary Materials for details.
To the best of our knowledge, our work demonstrates for the first time the distribution of \SI{1550}{nm} four-qubit entangled state in long telecom fibres, proving the viability of polarisation-encoded photons to remain highly entangled over long distances.

We also include the adjusted conference key rates when we perform the protocol with actively switched measurement bases.
In our experiment, this is accomplished by rotating wave plates with motorised stages that are slow compared to the clock rate of our sources.
As such, this leads to a reduced overall rate as shown in Fig.~\ref{fig:Figure2} (see Materials and Methods for details).

\begin{figure}[t!]
  \begin{center}
  \includegraphics[width=86mm]{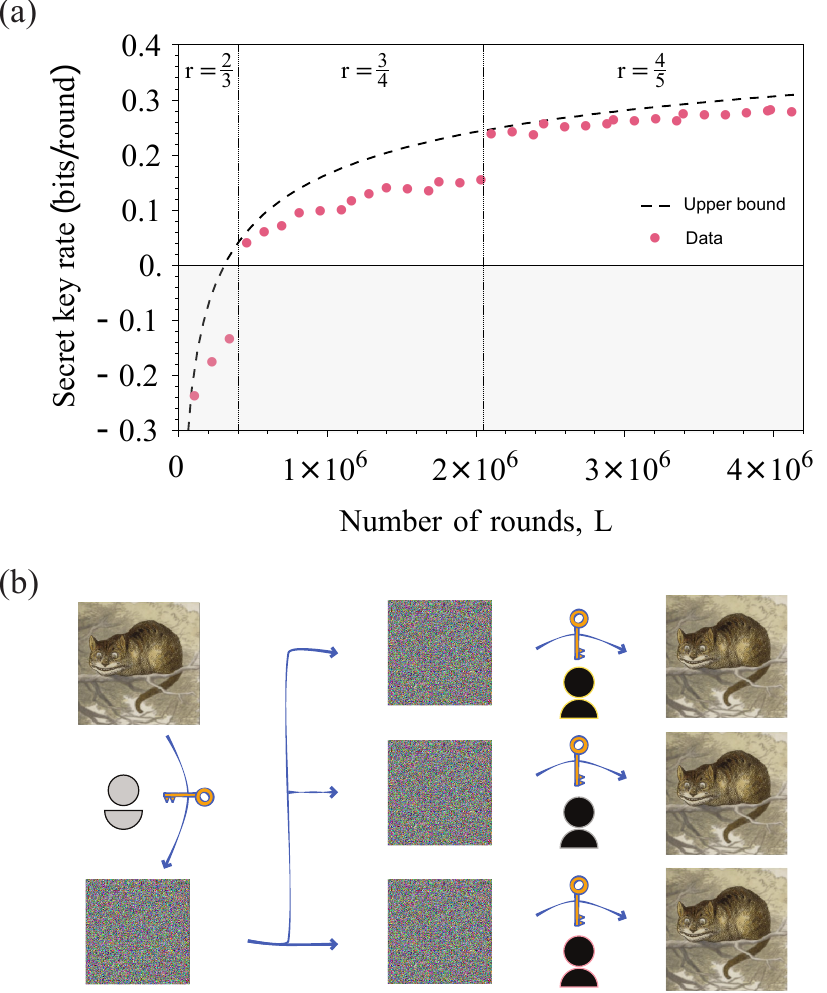}
  \end{center}
  \vspace{-2em}
  \caption{\textbf{(a)  Finite key results}. We implement all steps in the N-BB84 protocol for a range of \(L\) rounds to retrieve the final key of length \(\ell\) and evaluate the secret key rate, \(SKR = \ell / L\).
  In our experiment we employ LDPC codes with fixed code rates, \(r\), using the estimated QBER in each run.
  We implement privacy amplification using Toeplitz matrices, then remove a portion of the final key for the pre-shared bits used to encode the measurement-type rounds.
  The upper bound given by Eq.~\ref{eq_finite_key} is shown compared with the experimental data. 
  \textbf{(b) Encryption}. We generate an \(\epsilon_{tot}\)-secure conference key of \(1.15\times{10^6}\) bits.
  Using \(1.06\times{10^6}\) bits, Alice encrypts an image (8-bit RGB, 211 by 211 pixels) employing a one-time-pad-like scheme.
  Alice sends the encrypted image over a public channel allowing only Bob 1, Bob 2, and Bob 3, who share the conference key, to decode the image.}
  \label{fig:Figure3}
\end{figure}

The AKR results allowed us to establish upper bounds for several different fibre arrangements comparably quickly. 
To also show the N-BB84 performance in a real-world scenario, we implemented the complete protocol, including error correction and privacy amplification, for a fifth asymmetric fibre network \( \{5, 10, 20\} \) with a measured loss of \SI{9.53}{dB} in total. Due to the low rates, we need to apply finite-key analysis for this step, i.e., the secret key rate (SKR) is adjusted to account for finite statistics from parameter estimation.
For our experiment, we determine the optimal fraction of type-2 measurements to be $p=m/L=0.012$.
With this value of $p$, the amount of information reserved for the pre-shared key is $h(p)=0.093$ (see Materials and Methods for more details). Moreover, we set a total security parameter i.e., the maximal probability that an eavesdropper gains non-zero information about the key to be $1.8\times 10^{-8}$, see Supplementary Materials for details.

We obtain over \(4.09 \times 10^{6}\) type-1 rounds and \(5.01 \times 10^{4}\) type-2 rounds during 177 hours of continuous measurement.
Due to the long measurement time active polarisation feedback was implemented to minimise noise owing to thermal drifts in the laboratory (see Materials and Methods for details).
Once the raw key is distilled by all users, we implement one-way error correction using low-density parity-check (LDPC) codes complying with the Digital Video Broadcasting (DVB-S2) standard~\cite{morello2006dvb}.
The code was adapted to our multi-party scenario, simultaneously correcting Bob 1, Bob 2, and Bob 3 keys.
This step ensures that all parties share a common key, however it remains partly secret owing to information leaked during error correction, and any potential eavesdropping during the distribution step.
In order to reduce the information held by any potential eavesdropper, we implement one round of privacy amplification on the entire raw key, reducing its final length.
We use Toeplitz matrices for this purpose, a class of universal-2 hash functions~\cite{hayashi2011toeplitzUniversalityProof} that can be implemented efficiently for our given key size.

We estimate the theoretical performance of our post-processing steps by evaluating the noise parameters $Q_X=0.05$ and $\text{QBER}=0.0159$, which we use to calculate the upper bound set by Eq.~\ref{eq_finite_key} (see Materials and Methods) and plotted in Fig~\ref{fig:Figure3}(a) (dashed line).
When performing the protocol in earnest with a finite data set to estimate these parameters, we replace the Shannon limit for the error correction term $h(\text{QBER}^m + 2\xi_z)$ in Eq.~\ref{eq_finite_key} with the fraction of parity bits disclosed by Alice.

Finally, we use the secret conference key to encrypt an image of a Cheshire cat that is shared between the parties in a brief conference call (Fig.~\ref{fig:Figure3}b).
As shown, the key established by CKA enables any honest user in the group to share a secret message among all other honest parties.
This is in contrast with quantum secret sharing, a multi-user task demonstrated previously~\cite{Chen2005qss,Markham2008qss}, which requires co-operation among a majority subset of users to verify honesty and obtain the secret message.

\section*{Discussion}

A number of QCKA protocols have been proposed, including `N-six-state' with three  measurement bases~\cite{epping2017multi}. We implemented N-BB84 because it is experimentally friendly and enables higher rates for short keys~\cite{grasselli2018finite}. 
Novel QCKA variants include adaptations of two-party twin-field~\cite{Grasselli2019tfcka} and phase-matching~\cite{Zhao2020tfcka} protocols. These are attractive due to high rates achievable with weak coherent pulse sources. However, they require a common phase reference between all $N$ users which will be challenging in a network.

The N-BB84 protocol inherits several features from the entropic security proofs \cite{Tomamichel2012security} for the entanglement-based two-party protocols it is based on.
In particular, an eavesdropper's knowledge can be bound without full characterisation of all parties' measurement devices. The GHZ-state source can be completely untrusted.
Alice’s measurement device is trusted to ensure mutually unbiased measurement bases.
The Bob devices can then be untrusted since any deviation from ideal $X$ measurements negatively impacts the security parameter $Q_X$ ~\cite{grasselli2018finite}.
Finally, all measurement devices are assumed to be memoryless and detector efficiencies must be basis independent~\cite{Tomamichel2012security}.
Adapting the QCKA protocol for full (measurement-)device-independence is work in progress~\cite{holz2019mdiNQKDreply,ribeiro2019reply}. 

Another open question is that of the achievable rates in conference settings.
Experimental 2QKD key rates are bounded by the the well-known repeaterless bound~\cite{pirandola2017fundamental} established for point-to-point rates.
We remark that this bound does not apply to our scenario, where four users are connected to a common server according to some network topology. New bounds were recently found if repeaters are introduced in a chain-like network~\cite{pirandola2019end} showing that higher key-rates can in principle be achieved.
As our scenario omits repeaters these new bounds do not hold either, however we might expect similar improvements in the maximum key rates as opposed to standard end-to-end 2QKD protocols.
General bounds for distributing multi-partite entanglement in networks with nontrivial connectivity and noise have only very recently been established~\cite{takeoka2019asymmetricNetworks,Pirandola2019asymmetricNetworks,Pirandola2020asymmetricNetworks}.
For our own 4-user scenario we show in the Supplementary Materials that the QCKA rates have a non-trivial dependence on asymmetric network noise.
For direct GHZ-state transmission as demonstrated here, quantum CKA scales unfavourably with the number of users due to the exponential reduction in multi-photon detection due to unavoidable transmission losses.
However, loss will not be a problem in fully-featured quantum networks where CKA has a significant (N-1) rate advantage. 

The rate comparison between QCKA and 2QKD in Ref.~\cite{epping2017multi} did not account for the fact that 2QKD primitives incur not only post-processing overheads in respect to QCKA but also a cost on the secret key rates with respect to the underlying point-to-point rates.
In 2QKD, (N-1) unique pair-wise keys are transformed into a common secret key via bit-wise XOR operations.
If each bipartite key is \(\epsilon\)-secure then the final conference key is (N-1)\(\epsilon\)-secure owing to the composability of this multi-step approach~\cite{muller2009composability}.
To obtain an $\epsilon$-secure conference key, the individual keys have to be post-processed to a security threshold \(\frac{\epsilon}{N-1}\), which  lowers the final key rate.

Future experimental development will focus on increasing GHZ rates, the extension to more conference parties, and field tests in established fibre networks~\cite{dynes2019cambridge}. Multi-party entanglement applications beyond CKA include entanglement-assisted remote clock-synchronization~\cite{komar2014clocks}, quantum secret sharing~\cite{xiao2004emq,Chen2005qss,Markham2008qss}, and GHZ-based repeater protocols~\cite{kuzmin2019scalable}.

\section*{Materials and Methods}

\subsection*{Entangled photon source}
We produce photon-pairs using Type-II collinear spontaneous parametric down conversion (SPDC) implemented in a \SI{22}{mm} long periodically-poled KTP (PPKTP) crystal.
Both of our sources are optically pumped using a mode-locked laser operating with a nominal repetition rate of \SI{80}{MHz}, \SI{1.4}{ps} pulses and its central wavelength at \SI{774.9}{nm}.
A passive pulse interleaver is used to quadruple the \SI{80}{MHz} pulse train to \SI{320}{MHz}~\cite{broome2011reducing}.
The PPKTP crystals are embedded within a polarisation-based Sagnac interferometer ~\cite{Fedrizzi2007Sagnac} and pumped bidirectionally, using a half-wave plate to set diagonally-polarised light, to create polarisation-entangled photons at \SI{1549.8}{nm} in the approximate state:
\begin{equation}
\ket{\psi^{-}}=\frac{1}{\sqrt{2}}\left(\ket{h}\ket{v}-\ket{v}\ket{h}\right) ,
\label{psim}
\end{equation}
which we can map to any Bell state via local operation on one of the two photons.

With loose bandpass filters of \SI{3}{nm} bandwidth, we measure an average source brightness of $\sim 4100$ pairs/mW/s, with a symmetric heralding efficiency of $\sim 60\%$~\cite{graffitti2018design}. 
The average heralding efficiency reduces by \(\sim 12\%\) with a commensurate decrease of \(45\%\) in source brightness at the point of detection of the four users for zero fibre length.
We characterise each photon pair source by performing quantum state tomography, reconstructing the density matrix using a maximum-likelihood estimation followed by Monte-Carlo simulations based on Poissonian count statistics to determine errors.
For each source we obtain a typical two-photon Bell-state fidelity $F=95.58\pm0.15\%$ and purity $P = 92.07\pm0.27\%$, while entanglement is measured by concurrence $\mathcal{C}=92.38\pm0.21\%$.

The four-photon GHZ state is created by interfering one photon from each source on a polarising beamsplitter (PBS), which transmits horizontally and reflects vertically polarised photons.
Post-selecting on the case where one photon is emitted in each output, which occurs with a probability of $1/2$, we obtain the state
\begin{equation}
\ket{\text{GHZ}}=\frac{1}{\sqrt{2}}\left(\ket{hhhh}-\ket{vvvv}\right) \,, \label{ghz4}
\end{equation}
where $\ket{h}\equiv\ket{0}$ and $\ket{v}\equiv\ket{1}$ represent horizontal and vertical polarisations respectively.
We measure independent two-photon interference visibility of $92.96\pm0.95\%$ using \SI{100}{mW} pump power, and four-qubit state tomography returns a purity and fidelity of $P=81.39\pm0.83\%$ and $F=87.58\pm 0.48\%$ respectively.

\subsection*{Active switching}
Most QKD protocols require random switching of the measurement basis, either passively or actively, with each clock cycle. The same holds for the N-BB84 protocol, where users switch between the Z/X measurement bases according to a pre-agreed random sequence. Since all users implement the same measurement sequence, passive switching is not an option. 

As noted, \(p\) is typically small hence switching between bases occurs relatively infrequently. In addition the multi-photon detection rates in our experiment are low, hence the standard method of polarisation switching with electro-optic modulators would be excessive. We therefore implemented active switching using motorised rotation stages with switching speeds on the order of seconds---marginally slower than our average required switching periods, which reduces the maximum possible raw generation rate $g_R$.

We evaluate the adjusted generation rate $g'_R$ for the finite key scenario for the $\{5,10,20\}$ topology, by performing $1000$ rounds of the protocol with active basis switching. We set $p=0.02$, thus $20$ type-2 rounds are randomly allocated in the measurement sequence. We measured the reduced key generation rate and found $g'_R / g_R =  0.91$.

This adjustment ratio is rate dependent. We find the lower bound on $g'_R$ by assuming the type-2 rounds are never sequential hence each occurrence requires time to switch. This leads to the general expression,
\begin{equation}
    g'_R\geq\frac{1}{\tau_s p+\frac{1-p}{g_R}},
\end{equation}
where $\tau_s$ is the switching speed.
We use this equation to extrapolate the adjusted generation rates obtained in the asymptotic case as shown by orange dots in Fig.~\ref{fig:Figure2}.

\subsection*{Active polarisation control}
The optical fibre links in our experiment are realised by spools of bare SMF28 fibre.
Thermal drifts in the laboratory introduces unwanted rotations in polarisation which, if uncorrected, leads to added noise in the protocol.
These effects are typically negligible for short fibre lengths, e.g., in our testing we found the \SI{5}{km} spool added no observable noise greater than with a \SI{2}{m} fibre link, while the \SI{10}{km} and \SI{20}{km} spools showed significant added noise in \(Q_{\text{AB}i}\) measurements.

We implement active polarisation control to correct for these effects during key transmission to preserve low-noise operation throughout the protocol.
The feedback control loop is implemented by performing single-qubit tomography in each fibre to characterise the unitary transformation on the polarisation qubits.
We then use the polarisation optics in the measurement stages to undo the rotations on the qubits and perform measurements in the required basis.
In our setup we carry out one-qubit tomography of all four fibre links simultaneously, including post-processing, to obtain an estimate of the unitary operation and implement the corrective action on the motorised waveplates.
This takes less than 30 seconds and is performed once every \(\sim{20}\) minutes for an optimal tradeoff between maintaining a high duty-cycle while minimising bit error rates.

This feedback loop is not monitored for tampering by an eavesdropper. From a strict security perspective, a clever adversary may exploit this channel for executing a variant of the `time-shift' attack to gain control over a user's detectors. In principle, this can be mitigated by each user who swaps which detectors register the \(\{\ket{0}, \ket{1}, \ket{+}, \ket{-}\}\) events randomly in each round by rotating their waveplates. This can be performed locally without additional communication overhead among users.

\subsection*{Error correction using LDPC codes}
The use of LDPC codes allows one party to initialise the routine by computing $(j-k)$ parity-check bits from a block of $k$ raw bits using a $H_{(j-k)\times k}$ parity check matrix.
The ratio $r=k/j$ defines the code rate and higher code rates correspond to a smaller amount of information disclosed for error correction.
The DVB-S2 standard provides $H$ matrices already computed for a set of different code rates specified by an encoding block size of $j=64'800$ bits.
In our experiment, we set the code rate according to the estimated QBER using $m$ samples with appropriate \(\xi_Z\) correction.
From the provided set of code rates, we used $2/3$, $3/4$ and $4/5$ for small, mid and large values of L as shown in Fig.~\ref{fig:Figure3}(a).
Alice computes the parity-check bits by applying the parity check matrix $H$ to $k$-bit blocks of her raw key.
She then sends the parity-check bits, together with $H$, to all parties through authenticated classical channels.
With the information provided by the parity-check bits, each Bob implements a decoding algorithm on his respective raw key, consisting of simple addition, comparison and table look-up operations.
The codes used here have been modified from MatLAB communication packages based on the DVB-S2 standards~\cite{morello2006dvb}.
The number of parity bits communicated during EC is discarded to ensure security of the final conference key.

Optimal multi-user post-processing for QCKA is still an open question.
We know that CASCADE~\cite{brassard1993secret} can be more efficient than LDPC in the two-party setting for small error rates~\cite{elkouss2009efficient}.
However, as CASCADE relies on bidirectional communication, any benefits are quickly diminished by the increased communication overhead and required additional bit disclosures incurred between Alice and each Bob.
In contrast, LDPC codes disclose a fixed amount of information that depends only on the largest QBER between Alice and any of the Bobs in the network.
To the best of our knowledge, no proof exists for the optimal strategy to achieve the minimal bit disclosure rate when implementing error correction in multi-user QKD, and we leave this as an open question for future work.

\subsection*{Finite-key conference rate}
When using a finite number of rounds, the estimated parameters $Q_X^m$ and QBER from the m type-2 and type-1 rounds, are affected by statistical error which must be taken into account in the final key rate. The fractional key rate is given by,
\begin{equation}
\begin{split}
    \frac{\ell}{L} & =  \frac{n}{L}  [1-h(Q_X^m + 2\xi_X) \\
    & - h(\text{QBER}^m + 2\xi_Z)] - \log_2\left[\frac{2(N-1)}{\epsilon_{EC}}\right]^\frac{1}{L} \\
    & - 2\log_2\left[\frac{1-2(N-1)\epsilon_{PE}}{2\epsilon_{PA}}\right]^\frac{1}{L} - h(p) \,,
\end{split}
\label{eq_finite_key}
\end{equation}
where \(N\) is number of users in the protocol, \((\xi_X, \xi_Z)\) are finite-key correction terms and (\(\epsilon_{EC}, \epsilon_{PE}, \epsilon_{PA}\)) set the security parameters of our protocol, see Supplementary Materials for further details.
The final term in Eq.~\ref{eq_finite_key} is the portion of the final key removed after PA, to account for the preshared key used in marking the type-2 rounds.

\section*{Acknowledgements}
 This work was supported by the UK Engineering and Physical Sciences Research Council (grant number EP/N002962/1, EP/T001011/1). FG acknowledges financial support from the European Union’s Horizon 2020 research and innovation programme under the Marie Skłodowska-Curie grant agreement No 675662. MM acknowledges funding from the QuantERA ERA-NET Co-fund (FWF Project I3773-N36) and the UK EPSRC (EP/P024114/1).

\section*{Author contributions}
AF and MM conceived the project. MP, JH and PB performed the experiment and collected the data. JH and MP analysed the data. FG, MP and JH developed the theory results. All authors contributed to writing the manuscript.

\clearpage
\newpage
\renewcommand{\theequation}{S\arabic{equation}}
\renewcommand{\thefigure}{S\arabic{figure}}
\renewcommand{\thetable}{\Roman{table}}
\renewcommand{\thesection}{S\Roman{section}}
\setcounter{equation}{0}
\setcounter{figure}{0}

\section*{Supplementary Materials}

\subsection*{Experimental Noise}

As outlined in the main text, for the state employed in the protocol as in Eq.~\eqref{ghz4}, we expect $Q_X=0$ and $\text{QBER}=0$. However, in the experimental implementation, the values observed are always non-zero. In our setup as in Fig.~\ref{fig:Figure1}, the dominant sources of noise come from high-order generations in the PDC process and imperfect mode-matching at the PBS. A comprehensive model to account the effects of the noise on the expected value of $Q_X$ and QBER, is non-trivial and goes beyond the scope of this work. However, we provide some qualitative remarks and suggestions for improvements. 

First, we clarify that the high-order generations in the PDC process produce multi-photon events in an optical mode. In standard prepare-and-measure BB84 protocols this is vulnerable to the well-known photon-number-splitting attack. Here, an eavesdropper can perfectly remove a single encoded photon from the mode to learn the encoded key bit without detection. However, in entanglement-based QKD schemes like ours the eavesdropper cannot exploit multi-photon terms as the photons are not encoded in pure states, rather the encoding is derived from measurements performed by the users. Thus an eavesdropper cannot learn the key without adding noise. Nevertheless, the presence of these terms, even if unused by the eavesdropper, will introduce noise to the measurements and lower the overall secret key rate as we describe in the following.

\begin{figure}[b]
  \begin{center}
  \includegraphics[width=86mm]{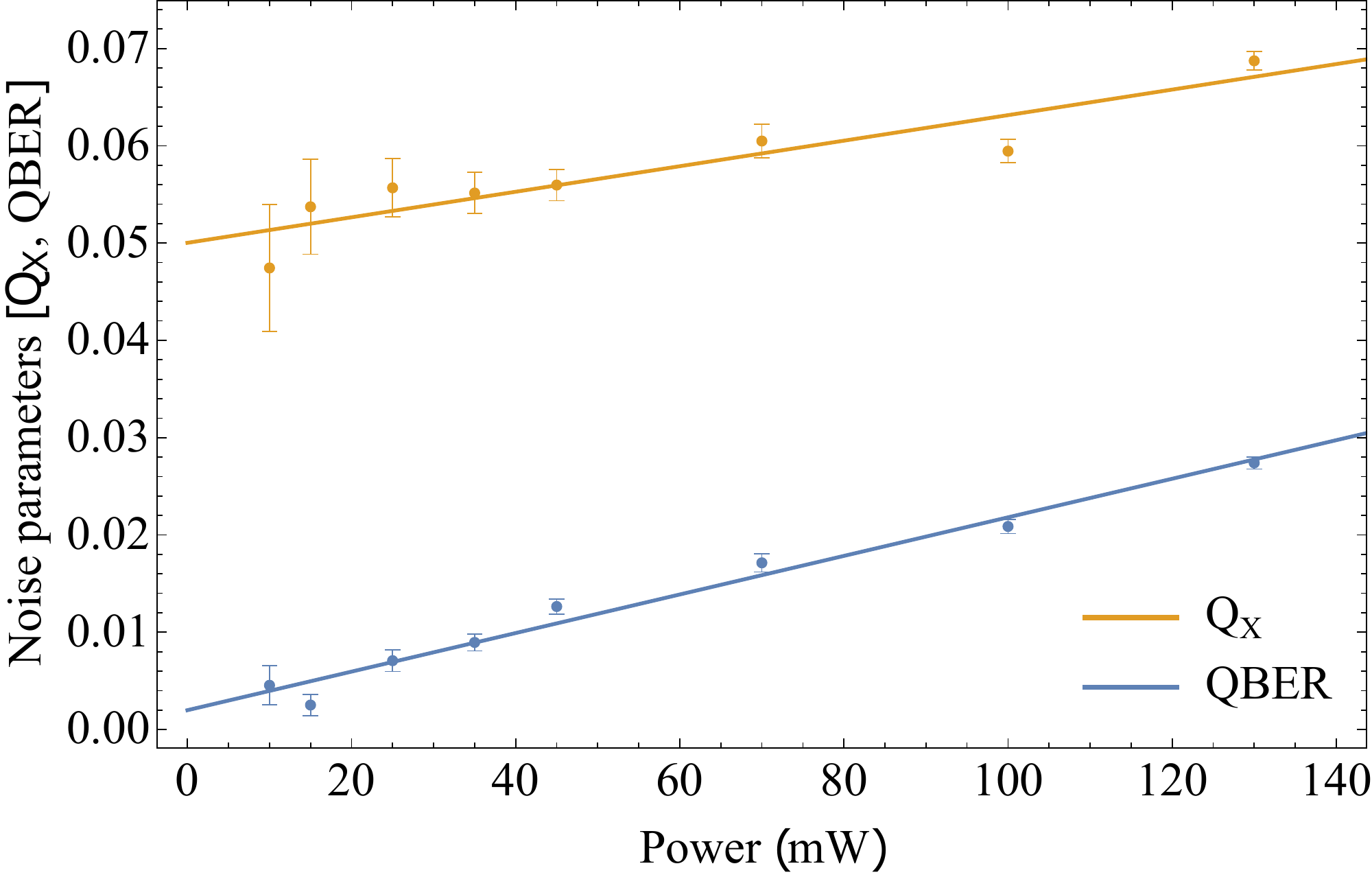}
  \end{center}
  \vspace{-2em}
\caption{$Q_X$ and QBER as a function of power. Within the range of power considered, the trend is linear although the slope for the QBER is greater than the slope for $Q_X$. Moreover, $Q_X$ is lower-bounded by the value of $0.05$ at zero-power.}
  \label{fig:FigureS2}
\end{figure}

Due to the probabilistic nature of the PDC process, there is always a non-zero probability that more than a single pair is generated within the crystal embedded in the Sagnac interferometer.
This effect can be quantitatively accounted for by the so-called signal-to-noise ratio (SNR), defined as the ratio of single-pair events over the multiple-pair events. Note that increasing the pump power decreases the SNR.
As shown in Fig.~\ref{fig:FigureS2}, both $Q_X$ and QBER depend indeed from the pump power.
The dependence is well fit by a linear trend, at least within the power range we considered.
Note that for the data shown represents the setup initially, which was later optimised for our experiments hence the values shown here are slightly greater than those reported in the main text at \SI{100}{mW}.
Importantly, whereas the QBER tends to $0$ in the limit of $\text{power}\rightarrow 0$, the $Q_X$ does not.
Qualitatively, this can be understood from the fact that the QBER only depends on the polarisation of the photons in the state in Eq.~\eqref{ghz4}, and not on their coherence.
On the other hand, the value of $Q_X$ is directly affected by the amount of coherence in the GHZ state considered for the protocol.
In turn, the state's coherence is influenced by all the degrees of freedom, i.e.; polarisation, photon-number, time and spectrum. Decreasing the power --therefore increasing the SNR-- only affects the purity in the photon-number degree of freedom but cannot affect the other degrees of freedom.
In particular, although our photons are spectrally filtered at the source, they retain some spectral mixture intrinsic to the PDC process.
This leads to non-ideal interference at the PBS, and therefore to a non-zero lower bound for the measurable $Q_X$.
Such lower bound can be linked to the experimentally measured visibility as following.
Assuming that the photons at the PBS successfully interfere with some probability $t$, we can write the state $\rho_o$ after the interference as:
\begin{equation}
    \rho_o = t\rho_{s} + (1-t)\rho_{f} .
\end{equation}
Where $\rho_s$ is the density matrix of the state in case of success, given by $\rho_s=\ket{GHZ}\bra{GHZ}$, and $\rho_f$ is the density matrix in case of failure given by $(\ket{hhhh}\bra{hhhh}+\ket{vvvv}\bra{vvvv})/2$. The expected $Q_X$ for this state is 
\begin{equation}
    Q_X = \frac{1 - \Tr[\rho_o \otimes X^{\otimes4}]}{2}=\frac{1 - t}{2}
\end{equation}
Note that for $t=1$, $Q_X=0$, and for $t=0$, $Q_X=1/2$.
Similarly, given the experimentally measured visibility $V_{\text{exp}}$ we expect $Q_X=0$ and $Q_X=1/2$ for $V_{\text{exp}}=1$ and $V_{\text{exp}}=0$ respectively.
We can thus, at least for these two extreme cases, interpret $t$ as $V_{\text{exp}}$.
Assuming that $t\approx V_{\text{exp}}$ in general, we have that for $V_{\text{exp}}=0.9$, $Q_X=0.05$ in accordance with our results (see main text).
It should be noted however, that the interference at the PBS is a coherent process, which might not be fully characterised by the simple model just presented.
Hence, in general, we can conclude that  $Q_X \gtrsim (1-V_{\text{exp}})/2$.

\subsection*{Security parameters in NBB84}

As stated in the main text, Eq.~\eqref{eq_finite_key} represents the achievable secret key rate of the NBB84 protocol when the parties perform a finite number of rounds $L$.
In other words, Alice needs to set the length of the PA output to Eq.~\eqref{eq_finite_key} in order to ensure that the established key is secure with security parameter $\epsilon_{tot}$. The security parameter $\epsilon_{tot}$ represents the maximal probability that a potential eavesdropper gains at least some information about the established key. It is related to the failure probabilities of the different stages of the protocol as follows: $\epsilon_{tot}= \epsilon_{EC}+\epsilon_{PA}+2\epsilon_{PE}$, where $\epsilon_{EC}$ is the maximal failure probability of the EC procedure and $\epsilon_{PA}$ represents the same in the case of PA, while the last term is related to the failure probability of the PE step.
In particular, the observed values $Q_{AB_i}^m$ and $Q^m_X$ in the $2m$ rounds devoted to PE might differ from the correspondent values $Q_{AB_i}^n$ and $Q^n_X$ characterizing the remaining $n=L-2m$ rounds which are used to extract the secret key. The deviation of $Q_{AB_i}^n$ and $Q^n_X$ is quantified by the theory of random sampling without replacement~\cite{bouman2010sampling}
and must be accounted for in the secret key rate Eq.~\eqref{eq_finite_key}, by taking the worst-case in order to preserve security. As shown in Ref.~\cite{grasselli2018finite}, the distance $|Q_{AB_i}^n-Q_{AB_i}^m|$ ($|Q^n_X-Q^m_X|$) between the pairwise bit discordance (the parameter $Q^n_X$) and its observed value is not larger than \(2\xi_Z\) (\(2\xi_X\)) with probability at least $1-\epsilon_Z$ ($1-\epsilon_X$), where:
\begin{align}
    \xi_{Z,X} = \sqrt{\frac{(n+m)(m+1)}{8nm^2} \ln \left(\frac{1}{\epsilon_{Z,X}}\right)}  \,\,.  \label{xi}
\end{align}
By combining the above statements one can deduce that:
\begin{align}
    &\Pr\left[Q^n_{X} \leq Q^m_{X} + 2\xi_X \,\,\wedge \,\, Q^n_{A B_i} \leq Q^m_{A B_i} + 2\xi_Z \,\,\forall i \right] \nonumber\\
	&\geq 1- \epsilon^2_{PE} \,\,,
\end{align}
where we defined the total PE failure probability $\epsilon^2_{\mathrm{PE}}$ as follows:
\begin{equation}
	\epsilon^2_{PE} \equiv (N-1)\epsilon_Z + \epsilon_X  \label{epsilon-PE} \,\,.
\end{equation}
Note that the probabilities $\epsilon_Z$ and $\epsilon_X$, and hence $\epsilon^2_{PE}$, can be chosen freely as to maximize the resulting secret key rate, with the only constraint that: $\epsilon_{PE} \leq \epsilon_{tot}$. Indeed, in our experiment we maximize the key rate in Eq.~\eqref{eq_finite_key} over the failure probabilities $\epsilon_{Z},\epsilon_{X},\epsilon_{EC}$ and $\epsilon_{PA}$ and over the fraction of type-2 rounds $p$, having fixed the security parameter to $\epsilon_{tot}=1.8 \times 10^{-8}$ and using preliminary estimations for QBER and $Q_X$. We obtain optimal values: $p=0.012$, \(\epsilon_{EC} \sim 10^{-13} \) and \(\epsilon_{PA} \sim 10^{-10} \). The optimal value for $p$ is then used to establish the fraction of type-2 measurements that need to be performed during data collection.
We remark that since $\epsilon_{EC}$ and $\epsilon_{PA}$ possess an operational meaning as described above, one needs to verify that the actual procedures implemented for EC and PA fail at most with probabilities $10^{-13}$ and $10^{-10}$, respectively. Due to the lack of a quantitative estimation of the failure probability characterizing the procedures adopted for EC and PA in our experiment, we could not verify that they are below the stated values. Nevertheless, we confirm that both procedures never failed in all the instances where they were used.

For further details, we refer the reader to Ref.~\cite{grasselli2018finite}.

\subsection*{Topology dependence in a conference key scenario}

Conversely from the standard Alice-Bob scenario, conference key protocols are performed over a network where different users are connected according to some topology, and each link might be some noisy quantum channel.
Therefore, in general, the conference key rates might depend on the noise distribution in the network opening a new problem absent in 2QKD.

Here, we study the 4-party network considered in the main text, i.e., four users connected to one common server, with the noise affecting each link modeled as a depolarising channel
\begin{equation}
    \mathcal{D}(\rho)=(1-\frac{3p}{4})\mathcal{I}+\frac{p}{3}(X\rho X +Y\rho Y + Z\rho Z) .
\end{equation}
Therefore, in general we can assume the channels of Alice, Bob 1, Bob 2 and Bob 3 to have noise parameters $p_\text{A}$, $p_{\text{B1}}$, $p_{\text{B2}}$, and $p_{\text{B3}}$, respectively.
For simplicity, we consider the case where Alice's channel is noiseless $p_\text{A}=0$ as the results are qualitatively the same.
In this case, the expressions of $Q_X$ and $Q_{ABi}$, for a depolarised 4-qubit GHZ state with noise parameters $p_{\text{B1}}$, $p_{\text{B2}}$ and  $p_{\text{B3}}$, are
\begin{align}
        &Q_X(p_{\text{B1}}, p_{\text{B2}}, p_{\text{B3}})=\frac{ \left[\left(p_{\text{B1}}-1\right) \left(p_{\text{B2}}-1\right) \left(p_{\text{B3}}-1\right)+1\right]}{2} \nonumber\\
    &Q_{ABi}(p_{\text{Bi}})=\frac{p_{\text{Bi}}}{2} \label{Qxdepol}
\end{align}
$Q_X$ depends on the noise parameters of all the channels, whereas $Q_{ABi}$ only depends locally on the noise parameter affecting the link connecting Alice and $\text{B}_i$.
Of course, both functions have a global minimum in $(p_{\text{B1}}$, $p_{\text{B2}}$, $p_{\text{B3}})=(0,0,0)$, that is when all the channels are noiseless.

\begin{figure}[t!]
  \begin{center}
  \includegraphics[width=86mm]{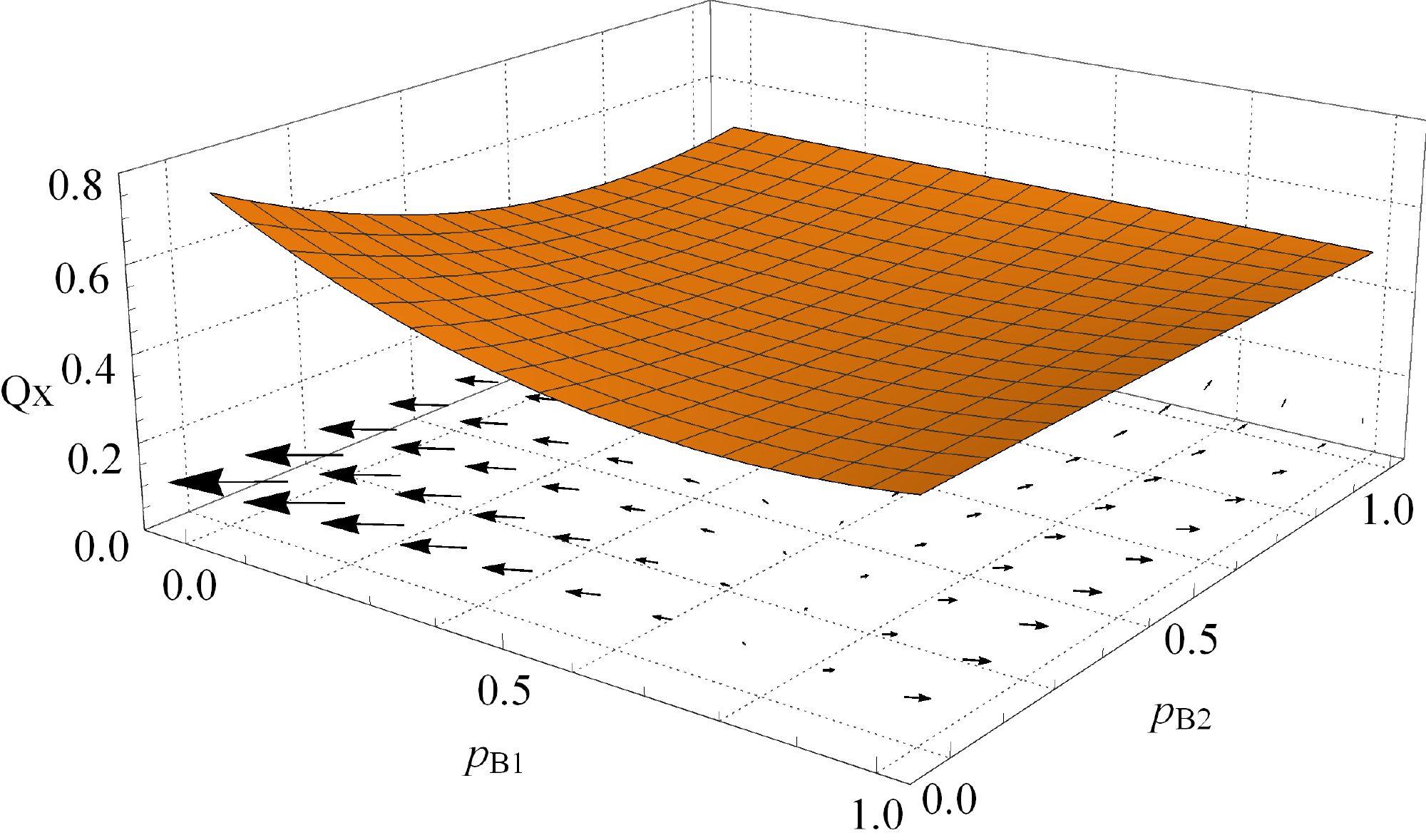}
  \end{center}
  \vspace{-2em}
\caption{Plot of $Q_X$ as a function of the noise parameters $p_{B1}$ and $p_{B2}$ characterising the depolarising channels (Eq.~\ref{Qxdepol}) of Bob 1 and Bob 2, respectively. The noise parameter of Bob 3 is fixed to: $p_{B3}=1.5-p_{B1}-p_{B2}$. We also insert the vector field of the gradient of $Q_X$ with respect to $p_{B1}$ and $p_{B2}$.}
  \label{fig:FigureS1}
\end{figure}

 What is interesting to study is whether both the functions have a minimum with the constraint $p_{\text{B1}} + p_{\text{B2}} + p_{\text{B3}}= c$ where $c$ is a constant in the interval $c\in[0,3]$. In practice, this corresponds to fix some total amount of noise strength $c$ on the network and finding which solution gives the highest key rate, i.e., the lowest $Q_X$ and $Q_{ABi}$. It is straightforward to see that the minimum of $\max_i{Q_{ABi}}$ is given by $p_{\text{B1}}=p_{\text{B2}}=p_{\text{B3}}= c/3$. To find the minimum of $Q_X$, we compute the gradient of $f(p_{\text{B1}},p_{\text{B2}})=Q_X(p_{\text{B1}}, p_{\text{B2}}, c - p_{\text{B1}}-p_{\text{B2}})$
\begin{align}
    &\frac{\partial f(p_{\text{B1}},p_{\text{B2}})}{\partial p_{\text{B1}}}= \frac{1}{2} (p_\text{B2}-1) (c-2 p_\text{B1}-p_\text{B2}))\\
    &\frac{\partial f(p_{\text{B1}},p_{\text{B2}})}{\partial p_{\text{B2}}}= \frac{1}{2} (p_\text{B1}-1) (c-p_\text{B1}-2 p_\text{B2}))
\end{align}
The plot in Fig.~\ref{fig:FigureS1} shows the function $f(p_{\text{B1}},p_{\text{B2}})$ for $c=1.5$ with at the bottom the vector field of the gradient as given by $\nabla f$. One can verify that the minimum of the function is in $p_{\text{B1}}=p_{\text{B2}}=p_{\text{B3}}= c/3$, therefore we conclude that the maximum conference key rate is achievable when the noise is symmetrically spread over the network. This result intuitively reflects the symmetry of the GHZ state, however in practice we can never assume the same amount of noise in all the channels. Nevertheless, it is important to note that the function is quite flat around the minimum. It follows that for small deviations from the symmetric configuration the effect on the key rate could be neglected. We leave open for investigation similar studies that account for different noise models.

\end{document}